\documentclass[aps,prev,twocolumn,preprintnumbers,floatfix,nofootinbib]{revtex4-1}

\usepackage{graphicx}
\usepackage{bm}
\usepackage{times}
\usepackage{slashed}
\usepackage{color}
\usepackage{color}
\def\br{\begin{eqnarray}}
\def\er{\end{eqnarray}}
\def\be{\begin{equation}}
\def\ee{\end{equation}}

\begin{document}

\title{Constraining the $Z^{\prime}$ Mass in 331 Models using Direct Dark Matter Detection}

\author{Stefano Profumo}
\author{Farinaldo S. Queiroz}

\affiliation{Department of Physics and Santa Cruz Institute for Particle Physics
University of California, Santa Cruz, CA 95064, USA}

\begin{abstract}
We investigate a so-called 331 extension of the Standard Model gauge sector which accommodates neutrino masses and where the lightest of the new neutral fermions in the theory is a viable particle dark matter candidate. In this model, processes mediated by the additional $Z^{\prime}$ gauge boson set both the dark matter relic abundance and the scattering cross section off of nuclei. We calculate with unprecedented accuracy the dark matter relic density, including the important effect of coannihilation across the heavy fermion sector, and show that indeed the candidate particle has the potential of having the observed dark matter density. We find that the recent LUX results put very stringent bounds on the mass of the extra gauge boson, $M_{Z^{\prime}} \gtrsim 2$~TeV, independently of the dark matter mass. We also comment on regime where our bounds on the $Z^{\prime}$ mass may apply to generic 331-like models, and on implications for LHC phenomenology. 
\end{abstract}
    
\maketitle

\section{Introduction}
\label{introduction}

The fundamental particle nature of the dark matter is one of the most pressing unanswered questions in science. The search for signals from dark matter that could shed light onto its particle nature is ongoing at a fast pace, and promises major breakthroughs on a very short time-scale. On the theory side, many dark matter candidates have been proposed and studied in detail, with a special role played by so-called WIMPs (an acronym for Weakly Interacting Massive Particles). WIMPs, which by definition possess a weak-interaction pair-annihilation cross section and a mass at the electroweak scale, naturally yield a thermal relic density consistent with the observed cosmological dark matter density (a fact sometimes indicated as ``WIMP miracle''). In addition, WIMPs are predicted to exist in many interesting particle physics models beyond the Standard Model (SM) such as the MSSM~\cite{mssm}, Left-Right Models~\cite{leftright}, Universal Extra Dimensions \cite{extradimension, Cornell:2014jza}, Little Higgs Models \cite{littlehiggs}, 331 models \cite{JCAP,331DM1,331DM2,331DM3}, and minimal extensions of the Standard Model (SM) \cite{minimalDM}. Less appealing dark matter candidates have been studied in Ref.\cite{Alves:2014kta}

In this paper, we focus on the dark matter phenomenology of a special class of theories, the so called 331 models, whose phenomenology has been studied in great detail from various particle physics standpoints, but not as far as dark matter searches are concerned. There exist many incarnations of 331 models in the literature, and many of them actually do not offer any viable dark matter candidate: these include the ``minimal'' 331 model \cite{331minimal}, the ``economical'' 331 model \cite{331econo}, and the 331 with two triplets of scalars \cite{331two}, among others \cite{amongothers}.  Supersymmetric \cite{331susyecono,331susyecono2} or Technicolor \cite{331tech} versions of these constructions might offer the prospect of having a viable dark matter candidate. However these supersymmetric and Techinicolor extensions have not yet addressed the issue of producing a suitable dark matter candidate in any detail. 

Concerning the minimal 331 models, in order to account for the dark matter, models must generically invoke an extended scalar or gauge sector, as pointed out in Ref.\cite{3311}. It is important to note that it has been claimed that the economical 331 model does feature a dark matter candidate, but a very severe fine-tuning is required in order to make the dark matter candidate stable. In particular one needs to invoke a very large suppression in the coupling $\lambda_3\sim 10^{-24}$ in the scalar potential in Eq.~(3.7) of Ref.\cite{DMeconomic}. Likewise, in Ref.\cite{331susyecono2}, the self-interacting dark matter scenario has been investigated. However, only the relic over-abundance requirement has been implemented so far. It would be interesting to investigate if this model has dark matter candidates with viable direct and indirect detection rates, and whether or not these rates are within reach of current experiments.

Here, we focus on the so-called 3-3-1LHN model, i.e. a model with  $SU(3)_c \otimes SU(3)_L \otimes U(1)_N$ gauge symmetry augmented with Left Handed heavy fermions. This model extends the SM by offering both 

(i) an elegant explanation to the observed neutrino masses, and 

(ii) a natural dark matter candidate, in marked difference from the other aforementioned 331 proposals. 

It has already been shown in Ref.\cite{331DM1, 331DM2} that this model may in principle feature two possible dark matter candidates, but that they cannot co-exist. Here, we consider the phenomenology of only one of these dark matter candidates, the lightest of the new, heavy fermions (which we indicate with $N$), with the purpose to determine the role of the $Z^{\prime}$ gauge boson as far as the dark matter phenomenology is concerned. 

In the present study we accurately calculate the dark matter thermal relic density, including new processes that have never been included in this context before (namely, coannihilation in the heavy fermion sector) and we derive stringent bounds on the mass of the $Z^{\prime}$ gauge boson by comparing the predicted scattering cross section off of nuclei with the most current limits from LUX \cite{LUX2013} and XENON100 \cite{XENON100}. These bounds we discuss here apply, {\it up to some extent}, to other extensions of the so called minimal 331 models in the sense that singlet neutral fermions are the most natural dark matter candidates in those models. 
In the latter setup the $Z^{\prime}$ would be the mediator and because the couplings of the $Z^{\prime}$ boson on those models are not so different from the model we investigate here our limits {\it do apply at some level}.

We also point out that our limits are complementary to other limits  on the $Z^{\prime}$ mass coming from colliders \cite{331collider1,331collider2} applicable to the model of interest, from FCNC \cite{331FCNC}, from oblique corrections to the STU parameters \cite{331STU},  and from muon decay \cite{331muon}. For complementary bounds on $Z^{\prime}$ gauge bosons in 331 models and simplified models see Ref.\cite{complementaryZpbounds}

The paper is organized as follows: In section \ref{sec1} we briefly introduce and review the particle content and key 3-3-1LHN model. In section \ref{darkmatter} we investigate the dark matter relic density in the model and we derive bounds on the mass of the $Z^{\prime}$ boson. Finally, we summarize and draw our conclusions in section \ref{sec:conclusions}. 

\section{The 3-3-1LHN Model}
\label{sec1}

We indicate with ``3-3-1 models''  extensions of the electroweak sector of the Standard Model where the electroweak sector $SU(2)_L \otimes U(1)_Y$ is enlarged to $SU(3)_L \otimes U(1)_N$. This extension is motivated by various, important problems not addressed by the SM, including the observed pattern of neutrino masses and mixing, the number of generations, as well as the existence of a suitable particle candidate for the dark matter. This model also reproduces the SM phenomenology as far as the Higgs sector is concerned, especially in light of recent experimental results, as shown for example in Ref.~\cite{331DM2}. For all these reasons,  3-3-1 models stand out as compelling extensions to the SM. 

The 3-3-1LHN we consider here has two noticeable distinct features compared to other incarnations of 3-3-1 models, namely: 

(i) the presence of heavy neutral fermions,  and 

(ii) the existence of two possible, distinct dark matter candidates. 

Below we briefly review the particle content and key features of the 3-3-1LHN model.


\subsection*{Leptonic Sector}

In the 3-3-1LHN model, leptons are arranged in triplet and singlet representations as follows:
\begin{eqnarray}
f_{aL} & = & \left (
\begin{array}{c}
\nu_{a} \\
e_{a} \\
N_{a}
\end{array}
\right )_L\sim(1\,,\,3\,,\,-1/3)\nonumber\\
     &e_{aR}& \sim(1,1,-1)\,,\, N_{aR}\,\sim(1,1,0),
\label{L}
\end{eqnarray}
where $a=1,2,3$ runs over the three lepton families, and $N_{a(L,R)}$ are new, heavy fermions added to the SM particle content. We emphasize that those heavy fermions (N) do not carry lepton number as we will clarify further.We will be hereafter using the above shorthand notation to refer to the quantum numbers of the symmetry group $SU(3)_c \otimes SU(3)_L\otimes U(1)_N$. For instance, as one can clearly see above, the leptons in the triplet are color singlets (1), triplets by $SU(3)_L$ (3) and have hypercharge $N=-1/3$, i.e $(1\,,\,3\,,\,-1/3)$.

\subsection*{Hadronic Sector}

The quarks in the theory, just like the leptons, come in triplets. In particular, the third generation lives in a triplet representation while the other two generations are in an anti-triplet representation of $SU_L(3)$, so that triangle anomalies cancel \cite{331minimal}. The corresponding quantum numbers are as follows:
\begin{eqnarray}
&&Q_{iL} = \left (
\begin{array}{c}
d_{i} \\
-u_{i} \\
q^{\prime}_{i}
\end{array}
\right )_L\sim(3\,,\,\bar{3}\,,\,0)\,, \nonumber \\
&&
u_{iR}\,\sim(3,1,2/3),\,\,\,
\,\,d_{iR}\,\sim(3,1,-1/3)\,,\,\,\,\, q^{\prime}_{iR}\,\sim(3,1,-1/3),\nonumber \\
&&Q_{3L} = \left (
\begin{array}{c}
u_{3} \\
d_{3} \\
q^{\prime}_{3}
\end{array}
\right )_L\sim(3\,,\,3\,,\,1/3)\,, \nonumber \\
&&
u_{3R}\,\sim(3,1,2/3),
\,\,d_{3R}\,\sim(3,1,-1/3)\,,\,q^{\prime}_{3R}\,\sim(3,1,2/3)
\label{quarks} 
\end{eqnarray}
where the index $i=1,2$ runs through the first two generations. The primed quarks $(q^{\prime})$ are new, heavy particles added to the SM particle content, with the usual fractional electric charges.

\subsection*{Scalar Content}
\label{scalarcontent}
The symmetry breaking pattern $ SU(3)_L\otimes U(1)_N \rightarrow SU(2)_L\otimes U(1)_Y$ $\rightarrow$ $U(1)_{QED}$ is reproduced with the introduction of three scalar triplets, namely
 
\begin{eqnarray}
\eta = \left (
\begin{array}{c}
\eta^0 \\
\eta^- \\
\eta^{\prime 0}
\end{array}
\right ),\,\rho = \left (
\begin{array}{c}
\rho^+ \\
\rho^0 \\
\rho^{\prime +}
\end{array}
\right ),\,
\chi = \left (
\begin{array}{c}
\chi^0 \\
\chi^{-} \\
\chi^{\prime 0}
\end{array}
\right )\,.
\label{scalarcont} 
\end{eqnarray}

The new scalars posses a general scalar potential of the form:

\begin{eqnarray} V(\eta,\rho,\chi)&=&\mu_\chi^2 \chi^2 +\mu_\eta^2\eta^2
+\mu_\rho^2\rho^2+\lambda_1\chi^4 +\lambda_2\eta^4
+\lambda_3\rho^4+ \nonumber \\
&&\lambda_4(\chi^{\dagger}\chi)(\eta^{\dagger}\eta)
+\lambda_5(\chi^{\dagger}\chi)(\rho^{\dagger}\rho)+\lambda_6
(\eta^{\dagger}\eta)(\rho^{\dagger}\rho)+ \nonumber \\
&&\lambda_7(\chi^{\dagger}\eta)(\eta^{\dagger}\chi)
+\lambda_8(\chi^{\dagger}\rho)(\rho^{\dagger}\chi)+\lambda_9
(\eta^{\dagger}\rho)(\rho^{\dagger}\eta) \nonumber \\
&&-\frac{f}{\sqrt{2}}\epsilon^{ijk}\eta_i \rho_j \chi_k +\mbox{H.c}.
\label{potential}
\end{eqnarray}
with $\eta$ and $\chi$ both transforming as $(1\,,\,3\,,\,-1/3)$
and $\rho$ transforming as $(1\,,\,3\,,\,2/3)$.

The scalar triplets above are introduced in order to generate masses for all fermions in the model
after the neutral scalars $\eta^0, \rho^0\ \mbox{and}\ \chi^{\prime 0}$ develop a vacuum expectation value different from zero.

\subsection*{Discrete Symmetry}
To ensure the stability of the theory's dark matter candidate, we invoke here a discrete symmetry quite similar to the R-parity of the minimal supersymmetric SM, which we indicate with $P=(-1)^{3(B-L)+2s}$, where $B$ is the baryon number, $L$ is the lepton number and $s$ is spin of the field. This symmetry commutes with the gauge symmetry and acts as follows:
\begin{eqnarray}
(N_L\,,\,N_R\,,\,d^{\prime}_i\,,\,u^{\prime}_3\,,\,\rho^{\prime +}\,,\,\eta^{\prime 0}\,,\,\chi^{0}\,,\,\chi^-\,,\, V^+\,,\,U^{0 \dagger}) \rightarrow -1,
\label{discretesymmetryI}
\end{eqnarray}where $d^{\prime}_i$ and $u^{\prime}_3$ are new heavy quarks predicted in the model due to the enlarged gauge group. The remaining fields all transform trivially under this symmetry. Note that the fermions N's do not carry lepton number. Therefore, the lightest neutral fermion odd under this parity symmetry is a possible dark matter candidate.

Additionally, we see that a particle which is a linear combination of the neutral scalars $\chi^{0}$ and $\eta^{\prime 0 *}$ might also be stable. We also note that the discrete symmetry also simplifies the mass spectrum of the model. In fact, Yukawa mass terms like $\bar{Q_{iL}}\chi^{*}d_{jR}, \bar{Q_{3L}}\chi u_{3R}$ and $\bar{Q_{iL}}\eta^{*} q^{\prime}_j$ among others, are forbidden in the Lagrangian, with significant simplifications in the resulting particle spectra. Such terms would for example induce mixing between the SM quarks and the new quarks $q^{\prime}$. 

Another possible way to guarantee the stability of our DM candidate would be by invoking the presence of an extra {\em gauge} symmetry which, after spontaneous symmetry breaking, would induce a residual unbroken $Z_2$ symmetry, as presented for instance in Ref.\cite{3311}. Here, however, we are not advocating that the 3-3-1 gauge symmetry is valid up to Planck scale. We could simply assume that such symmetry results from a more complex gauge group, for example such as the one proposed in Ref.\cite{331tech}, where a $Z_2$ symmetry arises as a result of spontaneous symmetry breaking of a gauge symmetry at high energy scales.

In summary, in the context of the 3-3-1LHN model there are two possible DM candidates: a complex scalar $\phi$ (the mass eigenstate resulting from the neutral scalar states in the theory) and a fermion $N_i$ (the lightest of the new heavy fermions). The most natural one, if all couplings in the theory are assumed to be of order one, is the fermion $N_1$ with normal mass hierarchy, and $N_3$ with an inverted hierarchy. We will hereafter assume a normal mass hierarchy, but the inverted hierarchy scenario, with $N_3$ as the lightest particle protected by the discrete symmetry, would not qualitatively be any different. In order to demonstrate that the $N_1$ is a good dark matter candidate we compute in detail below its thermal relic abundance and its scattering cross section off of nuclei, and compare our findings with current experimental bounds. 

\subsection*{Yukawa Sector}

As mentioned above, one of the benefits of introducing the symmetry of Eq.~(\ref{discretesymmetryI}) is to simplify the mass spectrum. The most generic Yukawa sector of the Lagrangian invariant under the 3-3-1 gauge and the G-symmetry is found to be
\begin{eqnarray}
&-&{\cal L}^Y =\alpha_{ij} \bar Q_{iL}\chi^* d^{\prime}_{jR} +f_{33} \bar Q_{3L}\chi u^{\prime}_{3R} + g_{ia}\bar Q_{iL}\eta^* d_{aR} \nonumber \\
&&+h_{3a} \bar Q_{3L}\eta u_{aR} +g_{3a}\bar Q_{3L}\rho d_{aR}+h_{ia}\bar Q_{iL}\rho^* u_{aR} \nonumber \\
&&+ G_{ab}\bar f_{aL} \rho e_{bR}+g^{\prime}_{ab}\bar{f}_{aL}\chi N_{bR}+ \mbox{h.c}., 
\label{yukawa}
\end{eqnarray}
where $\rho, \eta$ and $\chi$ are the scalar triplets introduced above.

One might notice that all fermions obtain Dirac masses, similarly to the Standard Model. The new fermions added to the SM, which will have Dirac mass terms as well, will have their masses proportional to the scale of symmetry breaking of the model. This model does not suffer from the problematic non-perturbative behavior at a few TeV that plagues minimal 331 models \cite{landaupole}, and hence one can easily push the scale of symmetry breaking up to very high energies. We will not consider this possibility here, however, since our goal here is only to derive bounds on the mass of the $Z^{\prime}$ boson based on direct detection searches of dark matter candidates at the electroweak scale.

\subsection*{Gauge Bosons}
\label{gaugebosons}

Due to the enlarged electroweak gauge group ($SU(2)_L \rightarrow SU(3)_L$) extra gauge bosons will be present in the 3-3-1LHC model, which we will indicate as $Z^{\prime}, V^{\pm},$ and $U^{0}$ and $U^{0\dagger}$. These bosons have masses proportional to the scale of symmetry breaking of the model, which are assumed here to be in the few TeV range. The charged currents involving these gauge bosons can be written as
\br
&&{\cal L}_{NH}=
-\frac{g}{\sqrt{2}}\left[\bar{\nu}^a_L\gamma^\mu e_L^a W^+_\mu +\bar{N}_L^a\gamma^\mu e_L^a V^+_\mu + \bar{\nu}^a_L\gamma^\mu N_L^a U^0_\mu  \right. \nonumber \\
&& \left. +\left(\bar{u}_{3L}\gamma^\mu d_{3L}  +\bar{u}_{iL}\gamma^\mu d_{iL}\right)W_\mu^+ +\left(\bar{q}^\prime_{3L}\gamma^\mu d_{3L}  +\bar{u}_{iL}\gamma^\mu q^\prime_{iL}\right)V_\mu^+  \right. \nonumber \\
&& \left. +\left(\bar{u}_{3L}\gamma^\mu q^\prime_{3L}  -\bar{q}^\prime_{iL}\gamma^\mu d_{iL}\right)U_\mu^0
+ {\mbox h.c.}
\right]\,,
\label{CC} 
\er
while the neutral current has the general form

\begin{eqnarray}
&&{\cal L}^{NC} =-\frac{g}{2 \cos\theta_W}\sum_{f} \Bigl[\bar
f\,  \gamma^\mu\ (g^\prime_V + g^\prime_A \gamma^5)f \, { Z_\mu^\prime}\Bigr],
\end{eqnarray}

\noindent
where $f$ are leptons and quarks, the couplings $g^\prime_V$ and $g^\prime_A$ 
are indicated in Tables \ref{tab2}, $g$ is the $SU(3)_L$  
coupling, and $\theta_W$ is the Weinberg angle. 

\begin{table}[t]
\begin{footnotesize}
\begin{center}
\begin{tabular}{|c|c|c|}
\hline
\multicolumn{3}{|c|}{$Z^{\prime}$ Interactions in the 331LHN } \\
\hline
Interaction &  $g^\prime_V$ & $g^\prime_A$   \\ 

\hline
$Z^{\prime}\ \bar u u,\bar c c  $ &
$\displaystyle{\frac{3-8\sin^2\theta_W}{{6\sqrt{3-4\sin^2\theta_W}}}}$  & 
$\displaystyle{-\frac{1}{2\sqrt{3-4\sin^2\theta_W}}}$  \\
\hline
$Z^{\prime}\ \bar t t$ & 
$\displaystyle{\frac{3+2\sin^2\theta_W}{{6\sqrt{3-4\sin^2\theta_W}}}}$  & 
$\displaystyle{-\frac{1-2\sin^2\theta_W}{2\sqrt{3-4\sin^2\theta_W}}}$  \\
\hline
$Z^{\prime}\ \bar d d,\bar s s  $ &
$\displaystyle{\frac{3- 2\sin^2\theta_W}{6\sqrt{3-4\sin^2\theta_W}}}$  & 
$\displaystyle{-\frac{{3-6\sin^2\theta_W}}{6\sqrt{3-4\sin^2\theta_W}}}$  \\
\hline
$Z^{\prime}\ \bar b b$ & 
$\displaystyle{\frac{3-4\sin^2\theta_W}{{6\sqrt{3-4\sin^2\theta_W}}}}$  & 
$\displaystyle{-\frac{1}{2\sqrt{3-4\sin^2\theta_W}}}$  \\
\hline
$Z^{\prime}\ \bar \ell \ell $ &
$\displaystyle{\frac{-1+4\sin^2\theta_W}{2\sqrt{3-4\sin^2\theta_W}}}$ &
$\displaystyle{\frac{1}{2\sqrt{3-4\sin^2\theta_W}}}$ \\
\hline
$Z^{\prime} \overline{N} N $ &
$\displaystyle{\frac{4\sqrt{3-4\sin^2\theta_W}}{9}}$ &
$\displaystyle{-\frac{4\sqrt{3-4\sin^2\theta_W}}{9}}$ \\
\hline
$Z^{\prime}\ \overline{\nu_{\ell}} \nu_{\ell} $ &
$\displaystyle{\frac{\sqrt{3-4\sin^2\theta_W}}{18}}$ &
$\displaystyle{-\frac{\sqrt{3-4\sin^2\theta_W}}{18}}$ \\
\hline
\end{tabular}
\end{center}
\end{footnotesize}
\caption{Coupling of the $Z^{\prime}$ with all fermions in the 3-3-1LHN model. Here $\theta_W$ is the Weinberg angle. It is worth pointing out that the interaction $Z^{\prime} \overline{N} N $ makes a crucial difference from previous 331 models proposals \cite{331minimal,331econo,331two}.}
\label{tab2}
\end{table}

The phenomenological aspects associated with the five gauge bosons in the model have been thoroughly explored in Ref.~\cite{gaugebosonsLHC}, to which we refer the interested Reader. The most striking phenomenological feature is the presence of charged gauge bosons. { At LEP-II charged gauge bosons with a light enough mass would have been produced in pairs via their photon and Z couplings. The production cross section depends only on the mass of the $V^{\pm}$ mass and and is large enough to rule out $M_{V^{\pm}} < \sqrt{s}/2 \sim 105$~GeV. 

At the LHC, $W^{\prime}$ bosons can be detected through
resonant pair production of fermions or electroweak bosons. The most commonly studied signal consists of a high-energy electron or muon and large missing transverse energy, with a peak in the number of events at $M_{W^{\prime}}/2$ as can be seen in Fig.1 of Ref.~\cite{Wprime1}. Assuming SM couplings with fermions, restrictive bounds were derived on the mass of the $W^{\prime}$, namely $M_{W^{\prime}} > 2.55$~TeV at $95\%$ C.L \cite{Wprime2}. However, this limit does not directly apply to our model for three reasons: 

(i) The boson $V^{\pm}$ couples, here, differently to the SM fermions, as one can clearly notice in Eq.~(\ref{CC}): some new particle from the 331 model is always present in the interactions involving the $V^{\pm}$ due to the parity symmetry; 

(ii) $V^{\pm}$ decays predominantly into WIMP plus electron ($N_1 e$) pairs; 

(iii) the production mechanism is not the same as for the $W^{\pm}$: in addition to Drell-Yan processes (photon and $Z$ s-channel mediated processes), there is a t-channel diagram mediated by new quark $q^{\prime}_1$, and three s-channel processes mediated by the Higgs, the scalar $S_2$ and the $Z^{\prime}$. 

In conclusion, one cannot straightforwardly apply the bounds found from ATLAS on $W^{\prime}$ mass to our model \footnote{Note that it is beyond the scope of this work to derive the precise impact of the LHC limits on this model. However, as aforementioned, these bounds are complementary to the ones we derive below.}.  LHC searches for the $W^{\prime}$ represent at some level a constraint on the mass of our charged gauge boson $V^{\pm}$, and they are complementary to the ones derived in this work using direct dark matter detection. A detailed study to translate bounds on the $W^{\prime}$ mass into a limit on the mass of the $V^{\pm}$ in our model is thus warranted in the future.}

\subsection*{Mass Eigenstates}

Spontaneous symmetry breaking in the present model is based on the non-trivial vacuum expectation value (vev) developed by the neutral scalars $\eta^0 ,\, \rho^0 ,\, \chi^{\prime 0}$. We indicate the vevs associated with each scalar as:
\begin{eqnarray}
 \eta^0 , \rho^0 , \chi^{\prime 0} \rightarrow  \frac{1}{\sqrt{2}} (v_{\eta ,\rho ,\chi^{\prime}} 
+R_{ \eta ,\rho ,\chi^{\prime}} +iI_{\eta ,\rho ,\chi^{\prime}})\,.
\label{vacua} 
\end{eqnarray}
There exist other neutral scalars in the spectrum, namely $\eta^{\prime 0}$ and $\chi^{0}$, which are enforced not to develop vevs in order to preserve the discrete symmetry, and therefore to guarantee the stability of our dark matter candidate. Notice in Eq.(\ref{scalarcont}) that $\rho^0$ and $\eta^0$ are $SU(2)$ doublets, therefore we expect $v_\eta$ and $v_\rho$ to be generically of the same order of magnitude. {In what follows, we  give analytical expressions for the particle spectrum utilizing, for the sake of simplifying the resulting expressions, the assumption $v_{\eta}=v_{\rho}=v=246/\sqrt{2}$~GeV; in our numerical study we have however computed all our results {\em without using any assumption on the vevs or on the constant couplings}. This is completely different from the simplifying assumptions used in previous works, such as Ref. \cite{JCAP,331DM1,331DM2,331DM3}}.

Once the pattern of symmetry breaking is established, one can straightforwardly obtain the ensuing mass eigenstates of the model. The SM fermion mass terms are unchanged, except for the neutrinos that acquire mass through dimension 5 effective operators \cite{331numasses}. We do not quote the resulting values for the neutrino masses, which can be made compatible with observation \cite{331numasses}, and we only exhibit a summary of the masses of the additional particles added to the SM below.

\begin{itemize}
\item {\bf Fermions}

The neutral fermions ($N_a$) shown in Eq.~(\ref{L}) are Dirac fermions with masses given by

\be
M_{N_a}=\frac{g^{\prime}_{aa}}{\sqrt{2}}v_{\chi^{\prime}}\,,
\label{mneut}
\ee
where $g^{\prime}_{aa}$ are the Yukawa couplings that appear in the last term of Eq.~(\ref{yukawa}). We assume all Yukawa couplings to be diagonal throughout this work.  

The three new quarks $q^{\prime}_a$ have their masses given by the first two terms of Eq.~(\ref{yukawa}) with,

\be
M_{q^{\prime}_a}=\frac{\alpha_{aa}}{\sqrt{2}}v_{\chi^{\prime}}\,.
\label{mquarks}
\ee

These new quarks do not play any role in the present analysis, and will be thus completely ignored from now on.  

\item {\bf Scalars}

After  spontaneous symmetry breaking the three CP-even neutral scalar mass eignestates ($H,S_1,S_2$) have masses
\begin{eqnarray}
M^{2}_{S_{1}} & = & \frac{v^{2}}{4}+2v_{\chi^\prime}^{2}\lambda_{1}\,, \nonumber \\
M^{2}_{S_{2}} & = & \frac{1}{2}(v_{\chi^\prime}^{2}+2v^{2}(\lambda_{2}+\lambda_{3}-\lambda_{6}))\,, \nonumber \\
M^{2}_{H} & = & v^{2}(\lambda_{2}+\lambda_{3}+\lambda_{6})\,.
\label{massashiggs}
\end{eqnarray}
$S_1$ and $S_2$ are new scalars particles added to the SM and have masses proportional to the scale of symmetry breaking of the model $v_{\chi^\prime}$, while $H$ is identified with the SM Higgs boson. The vev {\it v} which appears in Eq.~\ref{massashiggs} must be equal to $246/\sqrt{2}$~GeV, in order to reproduce the masses of the $Z$ and $W$ bosons. It has been shown in Ref.~\cite{331DM2} that the 3-3-1 Higgs boson $H$ reproduces the current results concerning the signal strength for the observation of the Higgs at the LHC. { We have fixed the sum $\lambda_{2}+\lambda_{3}+\lambda_{6}$ so that the Higgs mass of 125GeV is reproduced but we let the individual couplings free to vary in our numerical scan}.

Besides the three CP-even scalars, a new CP-odd scalar state ($P_1$) appears, with the following mass:
\begin{eqnarray}
M^{2}_{P_{1}} = \frac{1}{2}(v_{\chi^\prime}^{2}+\frac{v^{2}}{2}).
\label{massP1}
\end{eqnarray}

An additional complex neutral scalar also emerges which we indicate with $\phi$, with mass given by
\begin{eqnarray}
M_{\phi}^2 & = & \frac{(\lambda_{7} + \frac{1}{2} )}{2}[v^{2}+v_{\chi^\prime}^{2}].
\end{eqnarray}
 
Lastly, because of the presence of charged scalar fields in the triplet of scalars in Eq.~(\ref{scalarcont}), two massive charged scalars $h_1$ and $h_2$ arise, with masses
\begin{eqnarray}
M^{2}_{h^{-}_{1}} & = & \frac{\lambda_{8}+\frac{1}{2} }{2}(v^{2}+v_{\chi^\prime}^{2})\,, \nonumber \\
M^{2}_{h^{-}_{2}} & = & \frac{v_{\chi^\prime}^{2}}{2}+\lambda_{9}v^{2}\,.
\label{massash1h2}
\end{eqnarray}

Despite the fact that the 3-3-1LHN model has a large scalar content, none of these scalars will actually play a significant role in the phenomenology under scrutiny in this work\footnote{Note that as stated above we do not consider the possibility that the mass eigenstate $\phi$ be the lightest particle protected by the discrete symmetry and thus the model's dark matter candidate.}. We discuss them here primarily for the purpose of showing the richness of the mass spectrum predicted by this model. 

\item {\bf Gauge Bosons}

In the 3-3-1LHN model there is a total of 9 gauge bosons, arising because of the enlarged electroweak sector. Their masses are found to be, 
\begin{eqnarray}
M_{W^\pm}^2    &=& \frac{1}{4}g^2v^2\,,
\nonumber \\
M^{2}_{Z}      &=& m_{W^\pm}^2/c^{2}_{W}\,,
\nonumber \\
M^2_{V^\pm}    &=& m^2_{U^0} = \frac{1}{4}g^2(v_{\chi^\prime}^2+v^2)\,,
\label{massvec}
\end{eqnarray}and,

\begin{eqnarray}
&&M^2_{Z^\prime} = \frac{g^{2}}{4(3-4s_W^2)}[4c^{2}_{W}v_{\chi^\prime}^2 +\frac{v^{2}}{c^{2}_{W}}+\frac{v^{2}(1-2s^{2}_{W})^2}{c^{2}_{W}}].\nonumber\\
\label{Zprimemass}
\end{eqnarray}

It is important to emphasize that there are five gauge bosons in addition to the SM, which are within the reach of the LHC, since we assume that the corresponding masses, determined by the scale of symmetry breaking of the model, are in the few TeV range. Bounds on these particles' masses have been placed by the non-observation of certain classes of events \cite{331collider2,Wprime1,Wprime2}. In particular, a recent and restrictive limit was found on the mass of the $Z^{\prime}$ boson for the 3-3-1 model with right handed neutrinos using CMS data \cite{331collider2}, namely, $M_{Z^{\prime}} > 2.2$~TeV. This bound however, does not apply to our model, because the $Z^{\prime}$ here decays mostly into missing energy. For the regime where $M_{N_{a}}< M_{Z^{\prime}}/2$, the $Z^{\prime}$ decays mostly into neutral fermion pairs ($\overline{N_a}N_a$). Since we are assuming a normal hierarchy and $N_1$ is the DM candidate, the $Z^{\prime}$ will thus simply decay invisibly into dark matter particle pairs. Therefore, despite the production rate being the same, the branching ratio into charged leptons will be suppressed, and at some level the lower bound as well, as opposite to the 3-3-1 model with right handed neutrinos. Nevertheless, it is important to point out that in the mass regime where $Z^{\prime}$ boson cannot decay into the fermion pair the results found in Ref.~\cite{331collider2} do apply to our model. A variety limits have been placed on the mass of this boson and they come from different sources \cite{331FCNC,331STU,331muon} and from different models. In summary, the bounds derived here on the mass of this boson are complementary to those. 

We show in Fig.~\ref{widthZ1} how the mass of the $Z^{\prime}$ (in blue) and the total width (in red) vary with the scale of symmetry breaking of the model.  
\begin{figure}[!t]
\centering
\includegraphics[scale=0.6]{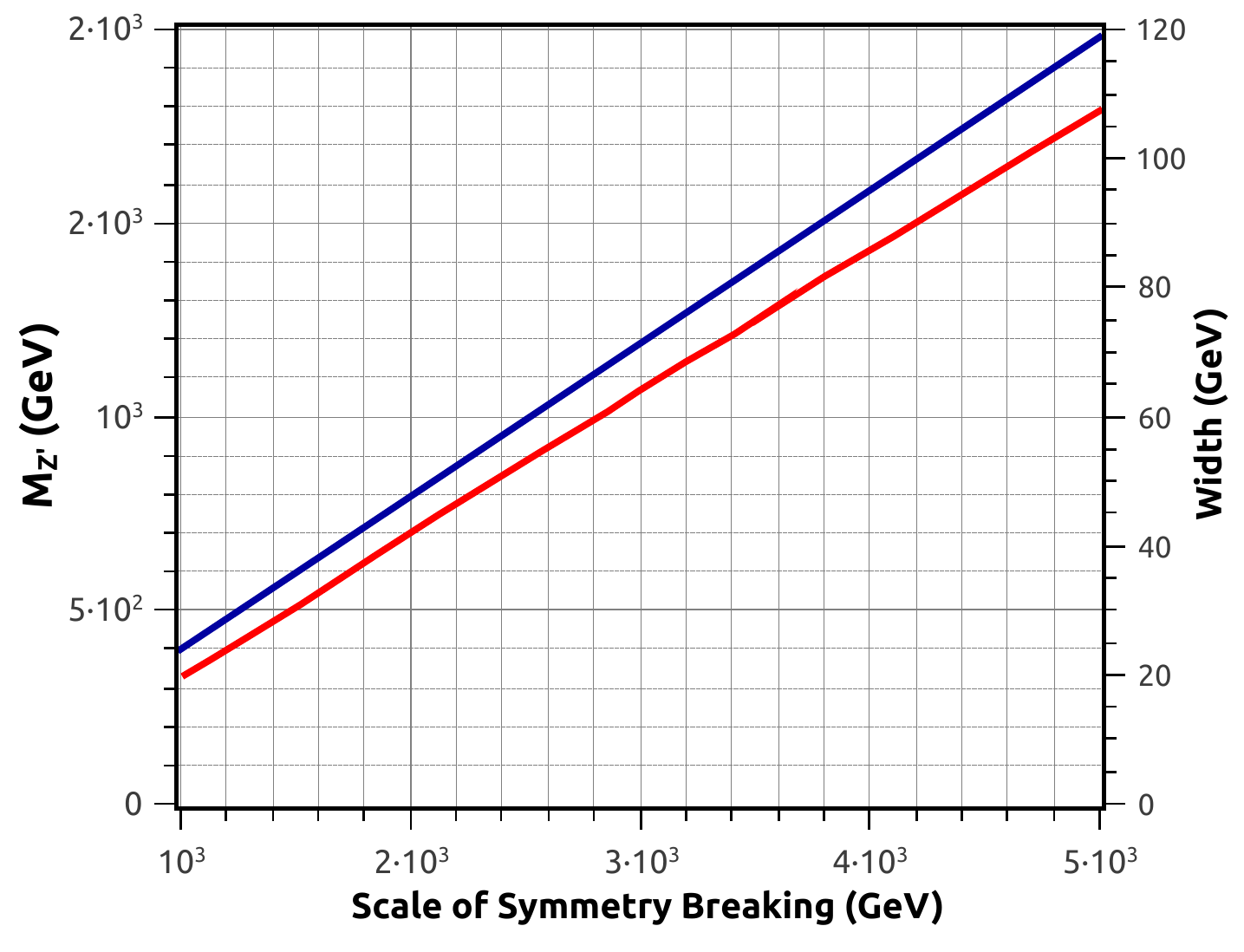}
\caption{Mass (blue) and total width (read) of the $Z^{\prime}$ as a function of the scale of symmetry breaking. }
\label{widthZ1}
\end{figure}
Since the mass of the $Z^{\prime}$ depends on the scale of symmetry breaking only, a bound on the mass of this boson translates into a limit on the whole mass spectrum of the model, because the masses of the new particles are all proportional to the scale of symmetry breaking. 
 
In summary, we have hereby briefly reviewed the key features of the 3-3-1LHN model. It will become clear from what follows that our results are complementary to other results, relevant for this class of models, obtained in the literature. We now turn to the phenomenology of the dark matter candidate, especially as a function of the $Z^{\prime}$ mass.
  
\end{itemize}

\section{Dark Matter}
\label{darkmatter}

\subsection{Thermal Relic Abundance}

\begin{figure}[!t]
\centering
\includegraphics[scale=0.6]{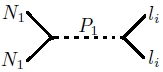}
\includegraphics[scale=0.6]{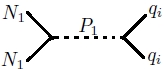}
\includegraphics[scale=0.6]{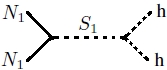}
\includegraphics[scale=0.6]{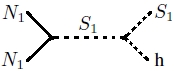}
\includegraphics[scale=0.6]{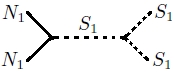}
\includegraphics[scale=0.6]{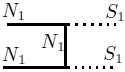}
\includegraphics[scale=0.6]{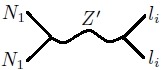}
\includegraphics[scale=0.6]{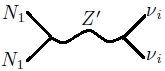}
\includegraphics[scale=0.6]{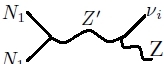}
\caption{Selected annihilation channels which contribute to the thermal relic density of our dark matter candidate $N_1$.}
\label{annhi}
\end{figure}

\begin{figure}[!t]
\centering
\includegraphics[scale=0.6]{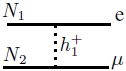}
\includegraphics[scale=0.6]{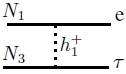}
\includegraphics[scale=0.6]{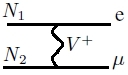}
\includegraphics[scale=0.6]{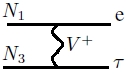}
\caption{Example co-annihilation channels which contribute to the abundance of our dark matter candidate $N_1$.}
\label{coanni}
\end{figure}

The calculation of the thermal relic abundance of our DM candidate (here assumed to be the heavy fermion $N_1$) in the 3-3-1LHN model follows standard techniques. To achieve the best possible numerical accuracy, we use a customized version of the micrOMEGAs package \cite{micromegas} on which we implemented the model of interest. In the present model, the thermal relic abundance is set by a wide variety of annihilation and co-annihilation processes, some of which are shown in Fig.~\ref{annhi} and \ref{coanni}, respectively.
%

It is important to notice that the new version of micrOMEGAs we employ includes the computation of 3- and 4-body final state processes. This is of great relevance in the present context, because it opens up new diagrams which had not been considered before, e.g., in Ref.~\cite{331DM1}-\cite{331DM2}. In addition to this, we include all relevant co-annihilation processes, such as those displayed in Fig.~\ref{coanni}, and we investigate the role of the gauge boson $Z^{\prime}$ in the overall abundance. In our calculations, we vary stochastically the mass splitting between our DM candidate $N_1$ and the heavier fermions $N_2$ and $N_3$ within $10\%$. We will see further that it is however the $Z^{\prime}$  gauge boson that plays the most important role in determining the abundance of the dark matter candidate and the associated direct detection rates.

\begin{figure}[!t]
\centering
\includegraphics[scale=0.6]{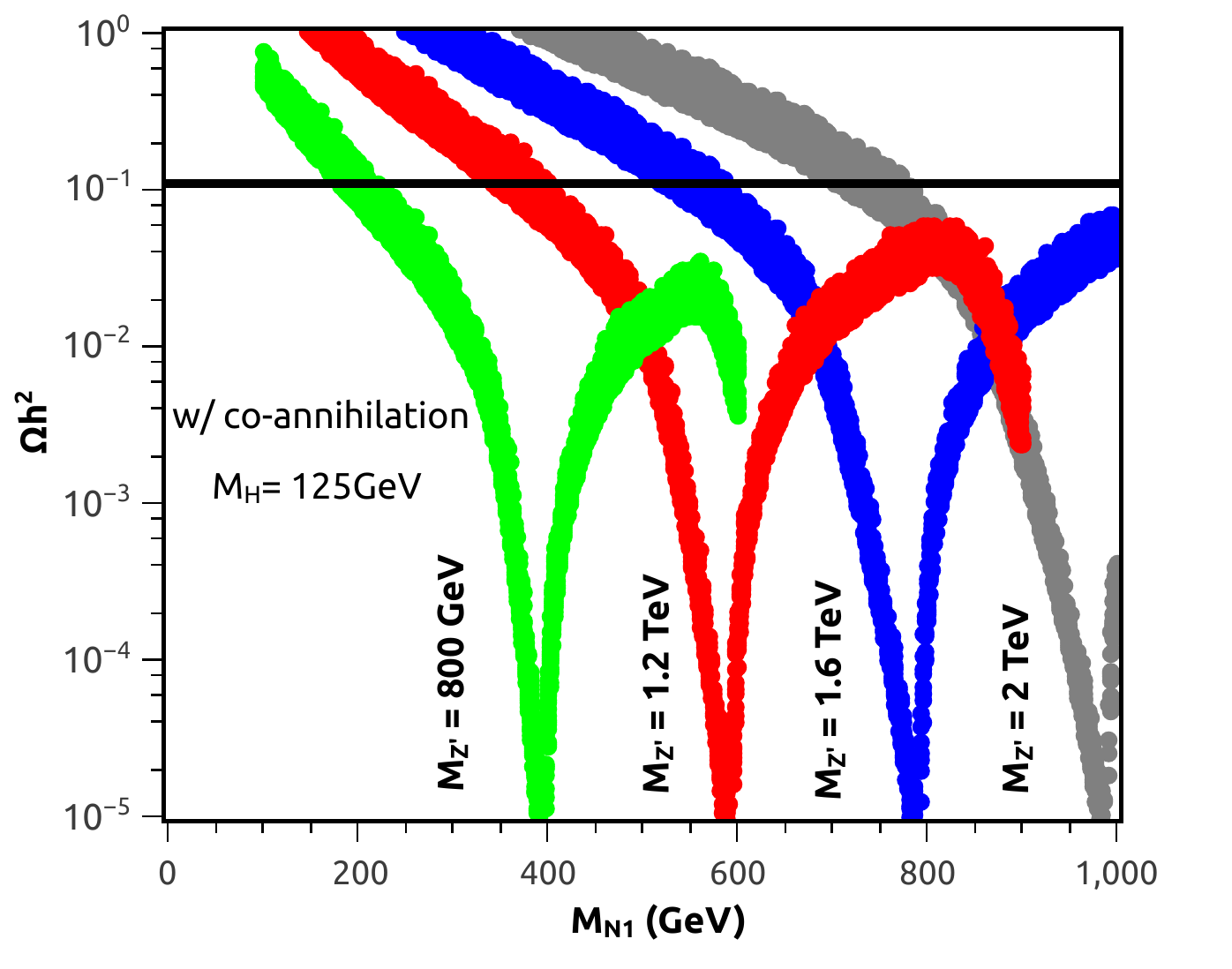}
\caption{Abundance of $N_1$ including co-annihilation as a function of its mass for four different value of the $Z^{\prime}$ mass. We can clearly notice a resonance at $M_{Z^{\prime}}/2$, indicating the major role played by the $Z^{\prime}$ in computing the abundance.}
\label{abunda2}
\end{figure}

{ In Fig.~\ref{abunda2} we show the abundance of the fermion $N_1$ as a function of mass, for four different values of the $Z^{\prime}$ mass when the co-annihilation processes are included. We keep the scale of symmetry breaking fixed, but we vary the masses of the particles. In particular, the masses of the heavier fermions $N_2$ and $N_3$ which, as stated above, are varied within 10\% of the $N_1$ mass. If we had kept the masses of the neutral heavy fermion far apart from each other no co-annihilation processes would be turned off. In the latter setup, we would obtain precisely the same curves shown in Fig.~\ref{abunda2} but the scatter points. In other words, we would have a fine line instead of a somewhat thick curve in Fig.~\ref{abunda2}.
 Throughout the parameter space of our model, we employ couplings of order one, and we use the values $v_{\chi^{\prime}}=2,3,4,5$~TeV while changing the mass of the WIMP. Taking all parameters of order one guarantees that all new particles lie at the $v_{\chi^{\prime}}$ scale, and enforces the DM candidate to be the $N_1$ (assuming fine-tuning in the $\lambda_7$ parameter, the scalar $\phi$ might become, in fact, lighter than $N_1$).\\
It is important to stress that the scalars $S_1,S_2$ and $P_1$ are irrelevant as far as the relic abundance of the neutral fermion is concerned, for the following reasons:

(i) In the scenario where $S_1$ is light no resonance rises. 

(ii) The pseudo-scalar $P_1$ induces velocity suppressed (p-wave) contributions to the abundance, and the annihilation cross section contribution is overwhelmed by the $Z^{\prime}$ one. We have explicitly investigated this scenario and found that in fact they are completely negligible; 

(iii) if one relaxes the usual assumption made in 331 models that $v_{\eta}=v_{\rho}$ and $f \sim V_{\chi}$, the conclusions do not change. We have also relaxed this assumption and found the same results. It is therefore clear that we can restrict our discussion to the contribution coming from the $Z^{\prime}$ gauge boson exclusivley, which we do hereafter}.

There are two important facts worth noting from the calculation of the $N_1$ thermal relic abundance. First, it is clear that the co-annihilation processes shown in Fig.~\ref{coanni} only produce some scatter in the abundance plot, which produces the ``thickness'' in the curves shown in the Fig.~\ref{abunda2}. Heavy fermions coannihilation processes, therefore, do not play a crucial role in setting the thermal relic abundance of the $N_1$ (our DM candidate). Second, the change in $v_{\chi^{\prime}}$ can be directly translated into a change in the $Z^{\prime}$ mass through Eq.~(\ref{Zprimemass}), in such way that the values for $v_{\chi^{\prime}}=2,3,4,5$~TeV effectively correspond to the choices $M_{Z^{\prime}}=0.8,1.2,1.6,2$~TeV. 

It is convenient to cast our results as a function of the $Z^\prime$ mass so we can clearly appreciate the effect of changing the $Z^{\prime}$ mass. For instance, for $v_{\chi^{\prime}}=2$~TeV ( $M_{Z^{\prime}}=0.8$~TeV), we observe that the thermal cross section has a resonance exactly at $M_{Z^{\prime}}/2=400$~GeV, and for this reason the resulting abundance is suppressed. This effect similarly appears for $M_{Z^{\prime}}=1.2,1.6,2$~TeV. This tells us that the $Z^{\prime}$ mediated processes in Figs.~\ref{annhi} are the most relevant ones, at least near resonance. In other words, by requiring the abundance of our WIMP to match observation, we can in principle constrain the mass of this gauge boson. However, one might notice from Figs.~\ref{abunda2}, that imposing the right abundance is not enough to obtain a bound on the $Z^{\prime}$ mass: for each value of $M_{Z^{\prime}}$ there is always a region of the parameter space, as small as it can be, that provides the right abundance. On the order hand, as we shall see in the next section, direct detection limits coming from LUX \cite{LUX2013}, rule out a large portion of the $Z^\prime$ mass range.


The figure also shows that, for a given value of the symmetry breaking scale, or equivalently of $M_{Z^\prime}$, the ``correct'' thermal relic density is always achieved at $M_{N_1}<M_{Z^\prime}/2$, and at $M_{N_1}>M_{Z^\prime}/2$ (on the ``other side'' of the resonance) only for massive enough $M_{N_1}\gtrsim$1 TeV. As $M_{N_1}\to M_{Z^\prime}$ the relic density drops again due to many additional coannihilation partners arising in the particle spectrum of the theory, and generically one gets a second viable value of $M_{N_1}$ (at fixed $M_{Z^\prime}$), but again only for massive enough $N_1$'s.


\subsection{Direct Detection and bounds on the $Z^\prime$}

In general, the WIMP scattering off of nuclei can be either spin-independent (SI) or spin-dependent (SD), depending or what sort of couplings are involved in the underlying theory. 
In our model, the dark matter candidate is a fermion that couples to quarks primarily through the $Z^{\prime}$ boson. This coupling results in a WIMP-Nucleon cross section that has both a SI and SD component. The SD WIMP-nucleon cross section is numerically larger than the SI WIMP-nucleon one. However, due to the well known enhancement from coherent scattering, the SI bounds on the WIMP-Nucleon cross section turn out to be stronger than the the SD ones. Therefore, we will limit our discussion to SI processes only. 

The differential event rate for elastic scattering of a WIMP with mass $M_{wimp}$ and a nucleus with mass $M_{nuc}$ is given by,
\be
\label{eq:dRdEr}
  \frac{dR}{dE_r}=\frac{N_T\,\rho_{DM}}{M_{wimp}}\int_{v_{min}}  v
  f_E(\vec{v}) \frac{d\sigma}{dE_r}(v,E_r)\, d^3\vec{v}\,,
\end{equation}where $N_T$ is the number of target nuclei per kilogram of the detector, $\rho_{DM}=0.3\ {\rm GeV/cm^3}$ is the local dark matter density, $\frac{d\sigma}{dE_r}(v,E_r)$ is the differential cross-section for the WIMP-Nucleus elastic scattering ,  $\vec{v}$ is the velocity of the WIMP relative to the Earth, $v_{min}$ is the minimum WIMP speed that can cause a recoil of energy $E_{R}$, and $f_E(\vec{v})$ is the the velocity distribution of the dark matter in the frame of the Earth (normalized to 1). This minimum velocity will depend on the energy threshold of the detector as well as on the masses of the WIMP and the nucleus.

In Eq.~(\ref{eq:dRdEr}) $dR/dE_r$ is the only measured quantity by direct detection experiments. The standard procedure is to plug in the Eq.~(\ref{eq:dRdEr}) the values of $N_T$ and $\rho_{DM}$, which are know quantities, and adopt some velocity distribution ($f_E(\vec{v})$), usually Maxwell-Botzmann, and assume some particular interaction between the WIMP and the nucleons, and the form factor, in such a way to determine $\frac{d\sigma}{dE_r}(v,E_r)$. 

The WIMP-Nucleus cross section is typically separated into a spin-independent (scalar) and a spin-dependent contribution as,
 
\begin{equation}
  \frac{d\sigma}{dE_r}= \left(\frac{d\sigma}{dE_r}\right)_{SI}
  +\left(\frac{d\sigma}{dE_r}\right)_{SD}\,,
\end{equation} but, as mentioned earlier, we will focus our attention to the SI only, since it provides stronger bounds. In this case the differential SI cross section might be written as,

\begin{equation}
\label{eq:dsigdEr}
  \frac{d\sigma}{dE_r}=\frac{M_{nuc}}{2\mu^2v^2} \sigma_0^{SI}
  F^2(q)\ ,
\end{equation}
where $q=\sqrt{2 M_{nuc} E_r}$ is the momentum transferred to the nucleus, $\sigma_0^{SI}$ is the SI cross sections at zero momentum transfer ($q=0$), $F^2(q)$ is the form factor that describes the dependence on the momentum transferred to the nucleus, in other words, it accounts for the coherence loss as the momentum transfer is increased.

Spin-independent contributions to the cross section may arise from scalar-scalar and vector-vector couplings in the Lagrangian:
\begin{equation}
  {\cal L}\supset \alpha_q^S
  \bar\chi\chi \bar qq +
  \alpha_q^V\bar\chi\gamma_\mu\chi\bar q\gamma^\mu q
  \,.
\end{equation}
The presence of these couplings depends on the particular particle physics model chosen for the dark matter candidate. In general one can write 
\begin{equation}
  \left(\frac{d\sigma}{dE_r}\right)_{SI}=\frac{M_{nuc} \sigma_0
  F^2(E_r) }{2\mu^2v^2} \,,
\end{equation}
where the nuclear form factor, $F^2(E_r)$, is the Fourier transform of the nuclear charge density and has the effect of suppressing the signal at large recoil energies, and $\sigma_0$ is the total WIMP-nucleon cross section, which has a scalar and vector component.

Scalar couplings lead to the following expression for the WIMP-nucleon cross section, 
\begin{equation}
  \sigma_0=\frac{4 \mu^2}{\pi} \left[Z f^p + (A-Z) f^n\right]^2 \,,
\end{equation}
with
\begin{equation}
  \frac{f^p}{m_p}=\sum_{q=u,d,s}\frac{\alpha^S_q}{m_q} f_{Tq}^p + 
  \frac{2}{27} f_{TG}^p\sum_{q=c,b,t}\frac{\alpha^S_q}{m_q} \,,
\end{equation}
where the quantities $f_{Tq}^{p}$ represent the contributions of the light quarks to the mass of the proton, and are defined as $m_pf_{Tq}^p\equiv\langle p|m_q\bar qq|p\rangle$. The second term is due to the 1-loop interaction WIMP-gluons through a colored loop diagram, with $f_{TG}^{p}=1-\sum_{q=u,d,s}f_{Tq}^{p}$. These quantities are related to the strange quark content in the nucleon and are determined from pion-nucleon scattering amplitude \cite{fornengo} and from baryon mass differences \cite{baryon}. 

The vector coupling is only present in the case of a Dirac fermion, such as our WIMP $N_1$.  The sea quarks and gluons do not contribute to the vector current. This means that only valence quarks contribute, leading to the following expression
\begin{equation}
  \sigma_0=\frac{\mu^2 B_N^2}{64\pi} \,,
\end{equation}
with 
\begin{equation}
  B_N\equiv \alpha_u^V(A+Z) + \alpha_d^V (2A-Z)
     \,.
\end{equation}

For a general WIMP particle with both scalar and vector interactions, the spin-independent contribution to the scattering cross section can be written as, 
\begin{equation}
   \left(\frac{d\sigma}{dE_r}\right)_{SI}=\frac{2\,m_N}{\pi
    v^2}\left[\left[Z f^p + (A-Z) f^n\right]^2 
     + \frac{B_N^2}{256}\right]
   F^2(E_r) \,.
\end{equation}
Most direct detection experiments choose to parametrize their results in terms of the scalar SI WIMP-nucleon cross section ($\sigma_{n}$ or $\sigma_{p}$), by rewriting the differential cross section as follows,
\be
\label{eq:SI}
   \left(\frac{d\sigma}{dE_r}\right)_{SI}=\frac{M_{nuc}\ \sigma_{i}}{2v^2\mu_n^2}\,\frac{\left[Z f^p + (A-Z) f^n\right]^2}{f_{i}^2}F^2(E_r) \,,
\ee
where
\be
  \sigma_{n,p}=\frac{4 \mu_{n,p}^2}{\pi}f_{n,p}^2\,,
\ee
where $\mu_{n,p}$ is the WIMP-nucleon reduced mass. In many cases the WIMP couples to neutrons and protons similarly, and in this situation $f^p \simeq f^n$, and therefore the scalar contribution can be approximated by 
\begin{equation}
\label{eq:crossSec}
   \left(\frac{d\sigma}{dE_r}\right)_{SI}=\frac{M_{nuc}\ \sigma_{n}\,A^2}{2v^2\mu_{n}^2}  
   F^2(E_r).
\end{equation}

Notice that for the vector coupling, the WIMP-Nucleus cross section would also scale with $A^2$ for $\alpha^V_u=\alpha^V_d$, and a similar definition for the WIMP-nucleon cross section would apply. Anyway, this $A^2$ enhancement typical for SI scatterings has lead many direct detection experiments to employ heavy targets such as Xenon and Iodine to boost the signal.  

\begin{figure}[!t]
\centering
\includegraphics[scale=0.5]{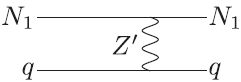}
\caption{WIMP-nucleon scattering process.}
\label{DDN1}
\end{figure}

We have thus far reviewed the procedure to calculate the SI WIMP-nucleon cross section determined by only one channel, shown in Fig.~\ref{DDN1}.
In theory there would exist two additional diagrams that could contribute to the WIMP-nucleon scattering. The second one is the 1-loop process with quarks running in the loop. However this process is not to relevant here because the fermion $N_1$ does not couple to the Higgs. The third, is a t-channel diagram mediated by the heavy pseudoscalar $P_1$ with mass given in Eq.~\ref{massP1}. Since the couplings involve a $\gamma_5$ matrix only, and the WIMP-nucleon scattering happens at the non-relativistic limit, this process is completely negligible. In any case, all processes are taken into account in the realization of the 3-3-1LHN model. Our results were obtained numerically using the micrOMEGAs package \cite{micromegas} and we let all coupling constants free to vary randomly.

\begin{figure}[!t]
\centering
\includegraphics[scale=0.55]{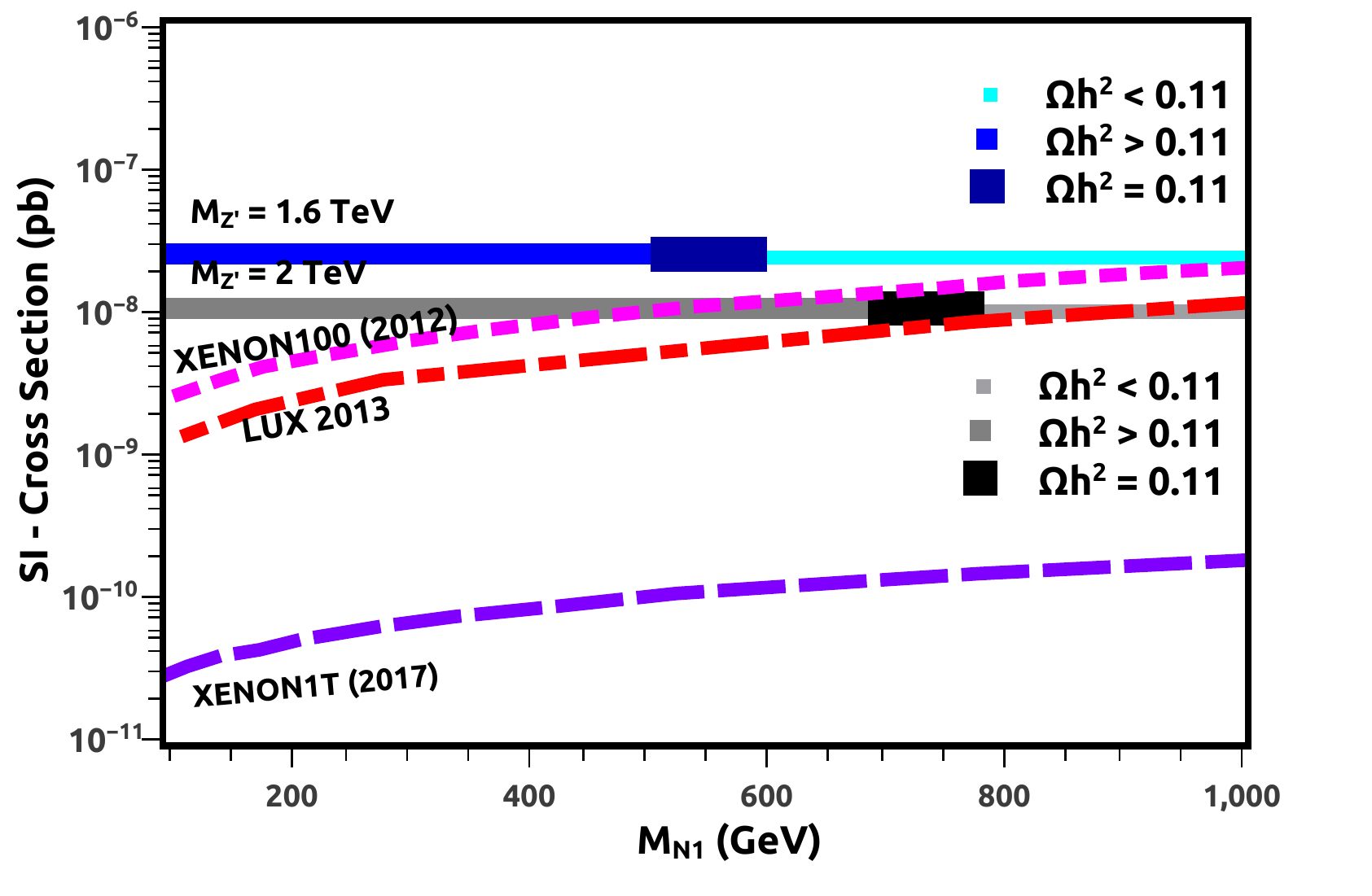}
\caption{SI scattering cross section off nuclei of the fermion $N_1$. See the text for details.}
\label{fig2}
\end{figure}

We summarize our numerical results for the $N_1$-nucleon scattering cross section as a function of the $N_1$ mass in Fig.~\ref{fig2}, for two values of the $Z^\prime$ mass. We set the symmetry breaking (vev) scale at $v_{\chi^{\prime}}=4$~TeV (green) and $v_{\chi^{\prime}}=5$~ TeV (blue). These values translate into $M_{Z^{\prime}}=1.6$~TeV and $M_{Z^{\prime}}=2$~TeV respectively, through Eq.~(\ref{Zprimemass}). Thicker lines indicate the $N_1$ mass range where a thermal relic density compatible with the observed dark matter abundance is achieved. The thick pink line indicates the XENON100 (2012) bound \cite{XENON100}: the region above the curve is excluded. The black dashed line indicates the anticipated 2017 XENON1T performance \cite{XENON1Tbound}, whereas the dashed red in the current LUX 2013 limit \cite{LUX2013}.

The figure shows that if one assumes that the $N_1$ is not heavier than $1$~TeV a lower bound $M_{Z^{\prime}} > 1.6$~TeV can be inferred from XENON100 data. However, if one assumes the $N_1$ to be much heavier than $1$~TeV this limit does not apply. For a $N_1$ lighter than $\sim 600$~GeV one will need a $Z^{\prime}$ much heavier than $2$~TeV in order to evade the XENON100 limits. However the recent LUX 2013 results literally excludes the whole parameter space with $M_{Z^{\prime}} < 2$~TeV. In other words, the direct detection data imposes a lower mass bound $M_{Z^{\prime}} > 2$~TeV. Also, it is apparent that there is only a very weak dependence on the $N_1$ mass. Since only gauge couplings are involved, the scattering cross section is determined by the mass of the WIMP and the $Z^{\prime}$ only. Consequently, the bound on the scattering cross section off nuclei can be converted into a limit on the mass of the $Z^{\prime}$ for a given WIMP mass, as we discuss below.

\begin{figure}[!t]
\centering
\includegraphics[scale=0.6]{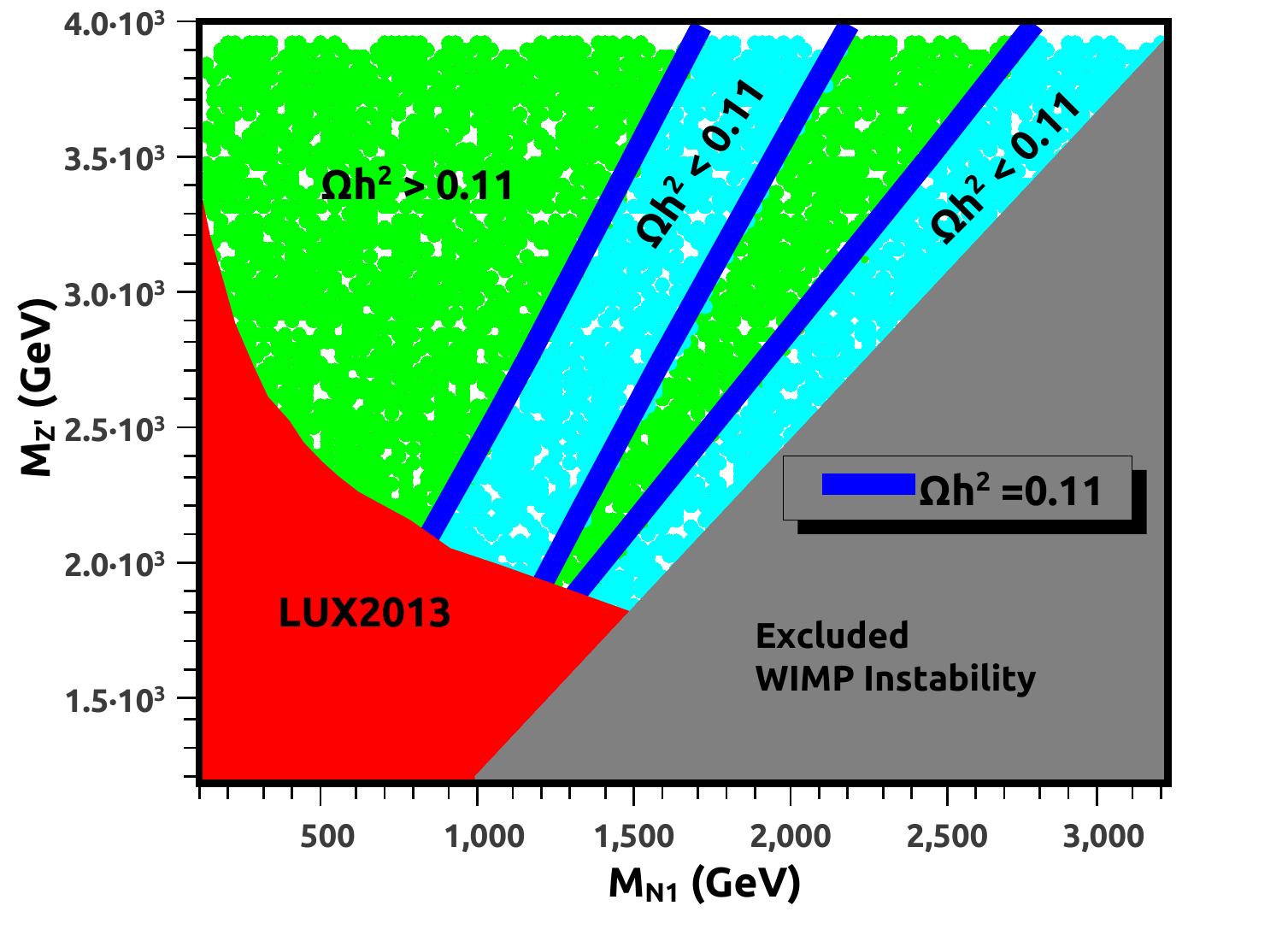}
\caption{The $M_{Z^{\prime}} , M_{N_1}$ parameter space. The red region is excluded by XENON100 bounds \cite{XENON100}. In the grey region the $N_1$ is not the DM candidate, and is thus unstable, allowing it to decay into $U^{0}\nu_e$, where $U^{0}$ is a neutral gauge boson and $\nu_e$ is the SM electron-neutrino according to Eq.~(\ref{CC}). The black, blue and green points indicate parameter space points where the thermal relic density $\Omega_{N_1}h^2 >0.11$, $=0.11$ and $<0.11$, respectively.}
\label{figZbound}
\end{figure}

The lower bound on the $Z^{\prime}$ thus depends on the $N_1$ mass regime we are considering. In Fig.~\ref{figZbound} we show the region of the parameter space ($M_{Z^{\prime}}, M_{N_1}$) which is allowed by direct detection searches of dark matter. The red region is excluded by LUX 2013 limits \cite{LUX2013}. The grey region is excluded because it induces the decay of $N_1$. In other words, $N_1$ is not the lightest particle charged under parity symmetry symmetry defined in Eq.~(\ref{discretesymmetryI}). For instance, when $M_{Z^{\prime}} = 1.2$~TeV, i.e. for $v_{\chi} = 3076$~GeV, the bosons $V^{\pm}$ and $U^{0}$ have masses close to $\sim 1000$~GeV, and because of the trilinear coupling involving theses boson and the fermion $N_1$, as one can see in Eq.~(\ref{CC}), the fermion $N_1$, which is assumed to be the dark matter candidate, cannot be heavier than about $1000$~GeV. For this reason the grey region reflects a $N_1$ stability requirement: if the $N_1$ is not the DM candidate it would not be stable. 

The figure also shows the regions where the $N_1$ thermal relic density is overabundant (green), under-abundant (light blue) and in accord (dark blue line) with the universal dark matter density. The structure of the relic density on the plane reflects what is shown in Figs.~\ref{abunda2}: the central funnel corresponds to the resonant annihilation mode via $Z^\prime$ exchange in an s-channel, while the right region, close to the instability region reflects the coannihilation with other particles in the theory (i.e. the right-most end of the curves in Fig.~\ref{abunda2}). These different regimes can be seen directly from Fig.~\ref{abunda2} and Fig.~\ref{fig2} as aforementioned.

As mentioned above, the bounds we discuss here apply, {\it at some level}, to other extensions of the so called minimal 331 models in the sense that singlet neutral fermions are the most natural dark matter candidates in those models. 
In the latter setup the $Z^{\prime}$ would be the mediator and therefore
our bounds would apply {\it up to some extent}, because the couplings of the $Z^{\prime}$ boson on those models are not so different from the model we investigate here. Moreover, these limits are complementary to other limits coming from colliders \cite{331collider2}, Flavor Changing Neutral Current processes \cite{331FCNC}, electroweak corrections to the S,T,U parameters \cite{331STU}, and from muon decay \cite{331muon}. More importantly, the limits on the mass of the $Z^{\prime}$ found here imply a bound on the scale of symmetry breaking that forces all particle masses to lie at a few TeV, if one considers all couplings to be of order one. 
As a final note, we warn the Reader that the limits we have derived here only apply under two assumptions:
\begin{itemize}
\item There is a discrete symmetry that guarantees the stability of our DM candidate ($N_1$) which arises from a spontaneous symmetry breaking of a gauge symmetry.

\item $N_1$ is the lightest particle charged under the discrete symmetry.
\end{itemize}

\section{Conclusions}\label{sec:conclusions}
In this paper we studied the phenomenology of the so-called 3-3-1LHN model. The model extends the weak interactions symmetry group from SU(2) to SU(3), it adds a variety of particles that fit in the new representations quarks and leptons belong to, and it adds a richer scalar sector, needed to obtain an appropriate pattern of symmetry breaking. In particular, the 3-3-1LHN model naturally encompasses heavy fermions and provides a viable dark matter candidate after imposing a suitable discrete symmetry.

While 3-3-1LHN models have been studied from a variety of particle physics standpoints, here we focused on the dark matter phenomenology. We implemented 3-3-1LHN models in a numerical code (micrOMEGAs) for the accurate calculation of the dark matter thermal relic abundance as well as the direct detection scattering rate. We then studied how direct detection results constrain the dark matter candidate mass and the mass of the $Z^\prime$, the latter in turn related to the scale of symmetry breaking of the model and to the mass of several other new particles in the theory. 

The thermal relic density of the dark matter candidate is set either by resonant annihilation through $Z^\prime$ exchange, or via coannihilation. We found that experimental direct detection results force the $Z^\prime$ mass to very large values if the WIMP mass is in the $\sim$TeV domain. In particular, we have outlined a lower bound, namely $M_{Z^{\prime}} \geq 2$~TeV. This mass value is in principle within reach of future LHC searches. Hence, in the next few years we expect either discovery or complementary bounds on the $Z^\prime$ boson of the 3-3-1LHN model. Either way, the LHC will shed light on the dark sector of this model. 

\acknowledgments

The authors would like to thank H.N. Long, Phung Van Dong, Yara Coutinho, William Shepherd, Patrick Draper, Carlos Pires and Paulo Rodrigues for their comments, and a special acknowledgement to Chris Kelso for his help on this work. We also thank Center for Theoretical Underground Physics and Related Areas (CETUP 2013-014) for their support and hospitality during the completion of this work. This work is partly supported by  the Department of Energy under contract DE-FG02-04ER41286 (SP,FSQ), and by the Brazilian National Counsel for Technological and Scientific Development (CNPq) (FQ).

\end{document}